\documentclass{Interspeech}

\interspeechcameraready 

\usepackage{amsmath,graphicx,hyperref}
\usepackage{booktabs}
\usepackage{amssymb}
\usepackage{multirow}
\usepackage{cite}

\title{Streaming T5-based Text-to-Speech Synthesis with Limited Lookahead}

\author[affiliation={1}]{Muyang}{Du}
\author[affiliation={2}]{Jason}{Roche}
\author[affiliation={1}]{Junjie}{Lai}

\affiliation{}{NVIDIA}{China}
\affiliation{}{NVIDIA}{USA}
\email{\{myrond,jroche,julienl\}@nvidia.com}

\begin{document}

\setlength{\abovedisplayskip}{3pt}
\setlength{\belowdisplayskip}{3pt}
\setlength{\abovedisplayshortskip}{2pt}
\setlength{\belowdisplayshortskip}{2pt}

\maketitle

\let\savethefootnote\thefootnote
\let\thefootnote\relax\footnotetext{Accepted at Interspeech 2026.}
\let\thefootnote\savethefootnote

\begin{abstract}
Streaming text-to-speech synthesis in cascaded LLM-TTS systems still faces latency challenges as most TTS models require full context before initiating generation. We present S5-TTS, a \textbf{S}treaming variant of T\textbf{5-TTS} that enables low-latency, word-by-word incremental speech synthesis through encoder-decoder language modeling and monotonic alignment learning. S5-TTS begins generating speech immediately after receiving the first few words, substantially reducing end-to-end response latency. To maintain quality under limited lookahead, we introduce a lookahead-causal masking mechanism with Conv-based auxiliary attention that preserves intelligibility and speaker similarity, and employ interleaved multi-source distillation to further restore naturalness. Experiments show that S5-TTS achieves comparable quality to full-context T5-TTS, supports zero-shot synthesis with high speaker similarity, and significantly reduces end-to-end latency for practical conversational AI systems.
~\footnote{\scriptsize Audio samples: \url{https://s5-tts.github.io/}}

\end{abstract}

\section{Introduction}
\label{sec:intro}

Large language models (LLMs) have become the cornerstone of generative AI, with state-of-the-art models adopting either decoder-only \cite{touvron2023llama, team2024gemma, yang2025qwen3} or encoder-decoder architectures \cite{chung2024scaling, zhang2025encoder, zhang2025t5gemma}. Recent advances in neural audio codecs \cite{zeghidour2021soundstream, kumar2023high} have enabled speech to be represented as discrete tokens, paving the way for LLM-based text-to-speech (TTS) models in both decoder-only \cite{kharitonov2023speak, chen2025neural, wang2025sparktts} and encoder-decoder forms \cite{ao2022speecht5, neekhara2024improving, battenberg-etal-2025-robust}. Compared to traditional TTS models \cite{lancucki2021fastpitch, kim2021conditional}, LLM-based TTS models benefit from large-scale training, resulting in significantly improved naturalness and diversity of generated speech. Unlike text-to-text modeling, where alignment is flexible, TTS typically exhibits near-monotonic alignment between text and speech. This property is difficult to capture with decoder-only models that rely solely on self-attention but can be effectively modeled through cross-attention in encoder-decoder architecture. For instance, T5-TTS \cite{neekhara2024improving} explicitly encourages monotonic alignment in decoder cross-attention, thereby effectively reducing hallucinations and enhancing robustness. Although recent Omni-modal models have emerged \cite{xie2024mini, fang-etal-2024-llama-omni, xu2025qwen2}, most interactive AI systems are still predominantly built upon a cascaded pipeline of text LLMs followed by TTS, due to its controllability and modularity. However, because most TTS models require full context and take full sentences as input before synthesis begins, these systems suffer from high end-to-end speech response latency.

Streaming incremental text-to-speech (iTTS) has been actively studied to reduce response latency before the full input sentence becomes available. Neural iTTS \cite{yanagita2019neural} demonstrates the feasibility of neural incremental synthesis by adapting Tacotron \cite{wang2017tacotron} with Griffin-Lim \cite{griffin1984signal} and shows that lookback acoustic context is crucial for natural prosody. Inspired by the prefix-to-prefix framework for simultaneous translation \cite{ma2019stacl}, subsequent work \cite{ma2020incremental} introduces the lookahead policy into Tacotron 2 \cite{shen2018natural} with Parallel WaveGAN \cite{yamamoto2020parallel}, showing that even a single future word substantially improves naturalness. Further analyses \cite{stephenson2020future} quantify this effect, reporting that one-word lookahead already recovers 88\% of the full-context representation, while two-word lookahead increases this to 94\%. Speech-T \cite{chen2021speech} extends the idea with a transducer-based architecture for unified ASR and TTS, likewise confirming that lookahead is sufficient to close the gap with non-streaming quality. More recently, InstantSpeech \cite{du2025instantspeech} reduces iTTS latency further with a fully parallel design inspired by FastPitch \cite{lancucki2021fastpitch} and Causal HiFi-GAN \cite{kong2020hifi}, while leveraging knowledge distillation to mitigate degradation under limited lookahead. Collectively, these studies establish a solid foundation for streaming iTTS and highlight the critical role of lookahead. Nevertheless, existing approaches remain restricted to single-speaker or few-speaker setups and show limitations in naturalness and zero-shot synthesis.

In this paper, we propose S5-TTS, a streaming variant of T5-TTS that enables low-latency, word-by-word incremental speech synthesis via encoder-decoder language modeling and monotonic alignment learning. Our key contributions are as follows: (1) S5-TTS can initiate speech synthesis immediately after the first few words become available, substantially reducing response latency in cascaded streaming LLM-TTS systems; (2) we introduce a lookahead-causal masking mechanism enabled by a Conv-based auxiliary attention, which allows the model to maintain high intelligibility and speaker similarity under limited lookahead; (3) we adopt an interleaved multi-source distillation strategy to effectively restore speech naturalness under limited lookahead; and (4) extensive experiments demonstrate that S5-TTS produces natural and intelligible speech comparable to the full-context T5-TTS, supports zero-shot synthesis for unseen speakers with high speaker similarity, and significantly reduces end-to-end speech response latency.

\section{Proposed Method}
\label{sec:proposed_method}

\subsection{Model Overview}

\begin{figure*}[tp]
    \centering
    \includegraphics[width=1.0\textwidth]{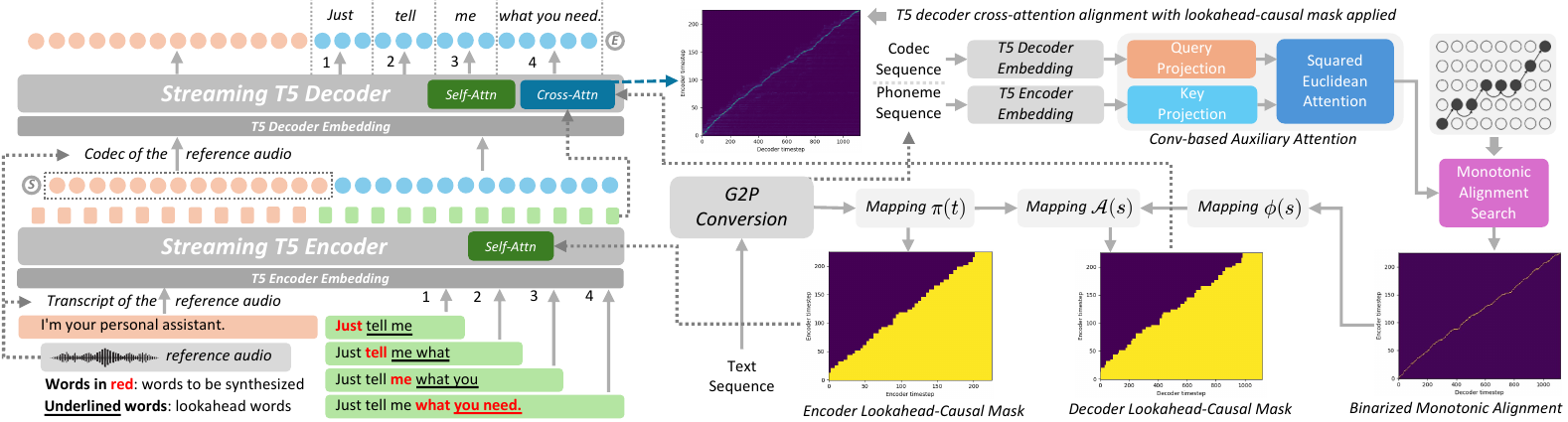}
    \caption{(Left) The overall architecture of S5-TTS. (Right) Lookahead-Causal Masks and Conv-based Auxiliary Attention.}
    \label{figs:architecture}
    \vspace{-6mm}
\end{figure*}

S5-TTS adopts a T5-based architecture, consisting of a parallel Transformer encoder and an autoregressive Transformer decoder. The encoder takes as input a phoneme sequence obtained via G2P conversion. At each decoding step, the decoder consumes the sum of the embeddings of the $K$ codec tokens predicted at the previous step and jointly predicts $K$ new codec tokens, each corresponding to one codebook in a Finite Scalar Quantization (FSQ) \cite{mentzer2023finite} audio codec. For each codebook, an embedding table is used, and its codec token is predicted using a dedicated linear projection head. As shown in Figure~\ref{figs:architecture}, S5-TTS is conditioned on reference audio during synthesis. Specifically, the transcript of the reference audio is concatenated with the target text as the encoder input, while the corresponding codec frames serve as a prompt to the decoder to guide generation.

Unlike T5-TTS, both the encoder and decoder in S5-TTS operate in a word-level streaming manner. For each word to be synthesized, the encoder processes the current word together with all preceding words and $k$ lookahead words, while the decoder autoregressively generates the codec chunk of the current word. During generation, the same lookahead setting is applied to G2P conversion for consistency, and we monitor the average cross-attention weights across all heads at each decoding step to identify word boundaries. Specifically, at decoder step $s$ while synthesizing word $i$, if the most attended encoder position shifts into the lookahead region, we consider the generation of word $i$ to be complete. The completion flag $c_s^i$ for word $i$ at decoding step $s$ is defined as:
\begin{equation}
c_{s}^{i} = \operatorname{argmax}(\alpha_{s}) > \Biggl[ \left( \sum_{j=1}^{i} W_j \right) - 1 \Biggr]
\end{equation}
where $\alpha_s$ denotes the cross-attention weight vector over all encoder steps at decoder step $s$, with the $\operatorname{argmax}$ index starting from 0, and $W_j$ representing the number of encoder steps corresponding to word $j$. When the final word of the target text enters the lookahead window, we stop monitoring the word completion flags and continue decoding until the decoder generates the final stop token. The codec chunk generated for word $i$ is converted into a waveform chunk by the FSQ codec decoder. To ensure smooth transitions between adjacent waveform chunks, we apply an overlap of two codec frames and perform crossfading using coefficients defined by a Hanning window.

To enable S5-TTS to operate under limited lookahead, we apply lookahead-causal masks to both the encoder and decoder, denoted as $M^{\mathrm{enc}}$ and $M^{\mathrm{dec}}$, respectively. The encoder mask $M^{\mathrm{enc}}$ restricts each word’s encoder steps to attend only to the current word, all preceding words, and $k$ lookahead words. Likewise, the decoder mask $M^{\mathrm{dec}}$ constrains each word’s decoder steps to cross-attend only to the corresponding encoder steps of the current word, all preceding words, and $k$ lookahead words. To ensure consistency between training and inference, both masks are applied in both phases. The construction of $M^{\mathrm{dec}}$ relies on an auxiliary attention module that is jointly trained with the encoder-decoder backbone using the auxiliary loss $\mathcal{L}_{\text{aux}}$. Detailed definitions of $M^{\mathrm{enc}}$, $M^{\mathrm{dec}}$, and $\mathcal{L}_{\text{aux}}$ are provided in the following subsections. Following T5-TTS, we additionally apply a Connectionist Temporal Classification (CTC) \cite{graves2006connectionist} loss to the cross-attention weight matrix $\alpha$, obtained by normalizing the cross-attention scores over encoder steps, to encourage monotonic alignments during training.

Formally, let $T$ and $S$ denote the numbers of encoder and decoder steps, respectively, and let each codebook have $V$ discrete codec tokens. The overall loss is defined as:
\begin{equation}
\begin{aligned}
\mathcal{L} &= \text{CE}\big(\text{softmax}(z), y\big) + \text{CTCLoss}(\alpha, \omega) + \mathcal{L}_{\text{aux}}
\end{aligned}
\end{equation}
where $z \in \mathbb{R}^{S \times Q \times V}$ denotes the decoder output logits, $y$ denotes the ground-truth codec frames, and $\omega = \{1, \dots, T\}$ denotes the ideal monotonic alignment path over encoder steps.

\subsection{Lookahead-Causal Masks}

Given a text with $N$ words, which may be either a full sequence (during training) or a partial sequence with lookahead words (during incremental inference), the encoder steps corresponding to the $n$-th word are denoted by the set $W_n$, where $n \in \{1,\dots,N\}$. Encoder steps are indexed by $t,t' \in \{1,\dots,T\}$, and decoder steps by $s \in \{1,\dots,S\}$.

\subsubsection{Encoder Mask Definition}

In the encoder, the mask is applied to the self-attention heads in all layers. Let $\pi(t)$ denote the mapping from encoder step $t$ to its corresponding word index. The encoder mask $M^{\mathrm{enc}} \in \{0,1\}^{T\times T}$ with $k$ lookahead words is defined as:
\begin{equation}
\resizebox{0.9\linewidth}{!}{$
M^{\mathrm{enc}}_{t,t'} =
\begin{cases}
1, & t' \in \bigl( \bigcup_{q=1}^{\pi(t)} W_q \bigr) \cup
\bigl( \bigcup_{q=\pi(t)+1}^{\min(\pi(t)+k,\, N)} W_q \bigr) \\
0, & \text{otherwise}
\end{cases}
$}
\end{equation}
where $\pi(t)$ can be directly obtained from the G2P conversion.

\subsubsection{Decoder Mask Definition}

In the decoder, the mask is applied to the cross-attention heads in all layers. Let $\mathcal{A}(s)$ denote the mapping from decoder step $s$ to its corresponding word index. The decoder mask $M^{\mathrm{dec}} \in \{0,1\}^{S\times T}$ with $k$ lookahead words is defined as:
\begin{equation}
\resizebox{0.9\linewidth}{!}{$
M^{\mathrm{dec}}_{s,t} =
\begin{cases}
1, & t \in \left( \bigcup_{q=1}^{\mathcal{A}(s)} W_q \right) \cup
\left( \bigcup_{q=\mathcal{A}(s)+1}^{\min(\mathcal{A}(s)+k, N)} W_q \right) \\
0, & \text{otherwise}
\end{cases}
$}
\end{equation}
where $\mathcal{A}(s)$ is obtained by first mapping decoder step $s$ to the corresponding encoder phoneme, and then to its word:
\begin{equation}
\mathcal{A}(s) = \pi(\phi(s))
\end{equation}
where $\phi(s)$ maps decoder step $s$ to its corresponding encoder step, representing the alignment between phonemes and codec frames. Although this alignment can be inferred from decoder cross-attention, it is required beforehand to construct the mask. We therefore predict $\phi(s)$ using an auxiliary attention module.

\subsubsection{Conv-based Auxiliary Attention}

Let $Q$ denote the decoder embedded codec frames and $K$ denote the encoder embedded phonemes. We first project them into a common attention space of dimension $C$ by $f_Q$ and $f_K$:
\begin{equation}
Q' = f_Q(Q), \quad K' = f_K(K)
\end{equation}
where both $f_Q$ and $f_K$ consist of a 3×3 convolution with ReLU activation followed by a 1×1 convolution.

The attention weights $e \in \mathbb{R}^{S \times T}$ between the $s$-th decoder step and the $t$-th encoder step are computed using the scaled negative squared Euclidean distance between projected $Q'$ and $K'$, and then normalized with a softmax over the encoder steps:
\begin{equation}
e_{s,t} = \operatorname{softmax}(-\delta \sum_{c=1}^{C} \left(Q'_{c,s} - K'_{c,t}\right)^2)
\end{equation}

Then the binarized alignment $\hat{A} \in \{0,1\}^{S \times T}$ can be obtained by taking the logarithm of the attention weights followed by Monotonic Alignment Search (MAS) \cite{kim2020glow}, defined as:
\begin{equation}
\hat{A}_{s,t} = \operatorname{MAS}(\log(e_{s,t}))
\end{equation}

Subsequently, the mapping $\phi(s)$ from decoder step $s$ to its corresponding encoder step is given by:
\begin{equation}
\phi(s) = \min \Big\{ t \,\big|\, \left(\sum_{i=1}^{t} \sum_{s'=1}^{S} \hat{A}_{s',i}\right) \ge s \Big\}
\end{equation}

During training, the auxiliary attention is jointly optimized with the S5-TTS backbone using the CTC loss, defined as:
\begin{equation}
\mathcal{L}_{\text{aux}} = \text{CTCLoss}(e, \omega)
\end{equation}

During inference, auxiliary attention is used only once to obtain the alignment of the reference audio, as $\mathcal{A}(s)$ for the generation phase can be recorded during incremental decoding by tracking the completion flags of each word.

\subsection{Interleaved Multi-Source Distillation}

Knowledge distillation has been shown to effectively improve the naturalness of lookahead-constrained TTS models \cite{du2025instantspeech}. In this work, we adopt a distillation strategy in which T5-TTS serves as the teacher and S5-TTS as the student. The pretrained student is distilled with supervision from both a paired text-audio dataset $\mathcal{D}_{\text{audio}} = \{(\textit{text}_i, \textit{audio}_i)\}$ and a text-only dataset $\mathcal{D}_{\text{text}} = \{\textit{text}_j\}$. For $\mathcal{D}_{\text{audio}}$, the teacher generates soft labels under decoder-side teacher forcing. For $\mathcal{D}_{\text{text}}$, soft labels are obtained via autoregressive decoding with sampling. To improve the reliability of supervision from text-only data, the decoded utterances are passed through an ASR filter, and only samples with zero WER are retained for distillation. During training, mini-batches from $\mathcal{D}_{\text{audio}}$ and $\mathcal{D}_{\text{text}}$ are interleaved within each gradient accumulation cycle, allowing their gradients to be jointly aggregated before each parameter update. We refer to this strategy as Interleaved Multi-Source Distillation (IMSD).

The distillation objective comprises three terms: an MSE loss on the final decoder hidden states $h$, a KL divergence loss on the decoder logits $z$, and a standard cross-entropy (CE) loss. The overall distillation loss is defined as:
\begin{equation}
\begin{aligned}
\mathcal{L}_{\mathrm{distill}} &= \lambda_h\,\text{MSE}(h^{\mathrm{stu}},h^{\mathrm{tea}})
+ \lambda_z\,\text{KL}(z^{\mathrm{stu}} \Vert z^{\mathrm{tea}}) \\
&\quad + \text{CE}\big(\text{softmax}(z^{\mathrm{stu}}), y\big)
\end{aligned}
\end{equation}
where $y$ is the ground-truth codec frames for $x \in \mathcal{D}_{\text{audio}}$ and the teacher-decoded codec frames for $x \in \mathcal{D}_{\text{text}}$ after ASR filtering.

\section{Experiments}
\label{sec:experiments}
\vspace{-1mm}
\subsection{Experimental Setup}
\label{experimental_setup}

\noindent$\blacksquare$\ \textbf{Datasets.}
For initial training, both S5-TTS and T5-TTS are trained on the full training splits of LibriTTS \cite{zen2019libritts} and HiFiTTS \cite{bakhturina2021hi} speech datasets, comprising 845.04 hours of speech from 2,319 speakers. For distillation, we use the same speech datasets together with additional conversational text sampled from UltraChat-200k \cite{ding2023enhancing}. These sentences are synthesized by the T5-TTS teacher using randomly selected reference audio from the LibriTTS and HiFiTTS training sets, and subsequently filtered by the Parakeet-TDT \cite{xu2023efficient} 0.6B ASR model. This process yields 1.3 million utterances, corresponding to 3,827.50 hours of synthetic speech with soft labels for distillation.

\noindent$\blacksquare$\ \textbf{Model Architectures.}
S5-TTS comprises 4 encoder layers and 8 decoder layers, each with 8 attention heads. Both the encoder and decoder use 768-dimensional embeddings, a feed-forward hidden size of 4096, and a dropout rate of 0.1, resulting in 160 million parameters. For T5-TTS, we adopt the same architecture described in its original paper. Both S5-TTS and T5-TTS employ the NeMo pretrained FSQ Mel codec model \cite{nvidia_mel_codec_22khz_medium} with 8 codebooks. This codec model encodes 22.05\,kHz audio at 86.1 tokens per second with 80 bits per token, corresponding to a bitrate of 6.9\,kbps. We use the phonemizer \cite{Bernard2021} for G2P conversion and adopt IPA as the phoneme set. In the auxiliary attention module, scaling factor $\delta$ is empirically set to $5\times10^{-5}$.

\noindent$\blacksquare$\ \textbf{Training and Evaluation.}
All models are trained on 4 NVIDIA B200 GPUs with a mini-batch size of 32 and a gradient accumulation factor of 4, resulting in an effective batch size of 128. Training is conducted for 250{,}000 steps using the AdamW \cite{loshchilovdecoupled} optimizer. The learning rate is linearly warmed up to $2\times10^{-4}$ over the first 10{,}000 steps, and then decayed to $1\times10^{-4}$ following a cosine schedule. For distillation, we finetune the pretrained model using a fixed learning rate of $1\times10^{-4}$ for 100{,}000 steps with $\lambda_h=10.0$ and $\lambda_z=1.0$. Training is performed in a purely language-modeling manner without reference audio. During inference, multinomial top-$k$ sampling is applied with $k=80$ and temperature 0.85. For evaluation, we primarily compare S5-TTS with T5-TTS, as both models adopt an encoder-decoder architecture and support zero-shot synthesis, allowing a more meaningful comparison than prior incremental TTS models that are restricted to single or few-speaker training settings. For comparison with other AR and NAR TTS models, we use their publicly released checkpoints. Subjective Mean Opinion Score (MOS) evaluation and preference tests are conducted on Prolific \cite{palan2018prolific} with 20 native English speakers.

\vspace{-1mm}
\subsection{Discussion}

\noindent\textbf{Analysis of Intelligibility and Speaker Similarity:}
We analyze the effect of the lookahead parameter \(k\) on S5-TTS by training three model variants with \(k = 1, 2,\) and \(3\), and comparing them to T5-TTS and ground-truth. Intelligibility is evaluated using WER and CER computed by the Parakeet-TDT 0.6B ASR model, while speaker similarity is measured as the cosine similarity between WavLM \cite{chen2022wavlm} embeddings of synthesized and ground-truth utterances. For both LibriTTS (including \emph{test-clean} and \emph{test-other}) and VCTK \cite{yamagishi2019vctk}, we randomly select 200 test utterances from unseen speakers for evaluation.

\begin{table}[h!]
\caption{Intelligibility, speaker similarity, and naturalness.}
\vspace{-3mm}
\centering
\small
\renewcommand{\arraystretch}{0.9}
\setlength{\aboverulesep}{0pt}
\setlength{\tabcolsep}{5pt}
\resizebox{1.0\columnwidth}{!}{%
\begin{tabular}{ccccccc}
\toprule
\textbf{Eval Set} & \textbf{Model} & \textbf{$k$} & \textbf{CER $\downarrow$} & \textbf{WER $\downarrow$} & \textbf{SSIM $\uparrow$} & \textbf{UTMOS $\uparrow$} \\
\midrule
\multirow{7}{*}{\shortstack{LibriTTS \\ \footnotesize (Unseen)}}
    & Ground Truth        & - & 0.91\% & 2.02\% & 1.0000 & \textbf{3.79} \\
    & T5-TTS    & - & 2.05\% & 3.20\% & 0.9356 & 3.77 \\
    \cmidrule(lr){2-7}
    & \multirow{3}{*}{S5-TTS}
        & 1 & 1.68\% & 3.03\% & \textbf{0.9376} & 3.63 \\
    &   & 2 & 2.12\% & 3.49\% & 0.9328 & 3.66 \\
    &   & 3 & 4.90\% & 7.27\% & 0.9335 & 3.58 \\
    \cmidrule(lr){2-7}
    & S5-TTS w/ IMSD  & 2 & \textbf{1.47\%} & \textbf{2.65\%} & 0.9340 & 3.72 \\
\midrule
\multirow{7}{*}{\shortstack{VCTK \\ \footnotesize (Unseen)}}
    & Ground Truth        & - & 0.56\% & 1.82\% & 1.0000 & 3.91 \\
    & T5-TTS    & - & 1.27\% & 1.63\% & \textbf{0.9379} & \textbf{3.99} \\
    \cmidrule(lr){2-7}
    & \multirow{3}{*}{S5-TTS}
        & 1 & 0.98\% & 1.51\% & 0.9350 & 3.88 \\
    &   & 2 & 1.09\% & 1.80\% & 0.9378 & 3.91 \\
    &   & 3 & 4.02\% & 5.70\% & 0.9291 & 3.86 \\
    \cmidrule(lr){2-7}
    & S5-TTS w/ IMSD  & 2 & \textbf{0.70\%} & \textbf{1.11\%} & 0.9362 & 3.93 \\
\midrule
\multirow{4}{*}{\shortstack{UltraChat \\ \footnotesize (Unseen)}}
    & T5-TTS    & - & 0.75\% & 1.06\% & 0.9776 & \textbf{4.28} \\
    \cmidrule(lr){2-7}
    & S5-TTS    & 2 & 0.29\% & 0.97\% & \textbf{0.9798} & 4.13 \\
    \cmidrule(lr){2-7}
    & S5-TTS w/ IMSD  & 2 & \textbf{0.27\%} & \textbf{0.83\%} & 0.9791 & 4.17 \\
\bottomrule
\end{tabular}
}
\label{tab:cer_wer_ssim}
\vspace{-6mm}
\end{table}

As shown in Table \ref{tab:cer_wer_ssim}, S5-TTS achieves intelligibility and speaker similarity comparable to T5-TTS when using one or two lookahead words. With \(k = 1\), S5-TTS even achieves lower WER than T5-TTS, which may be attributed to the limited lookahead encouraging the model to focus more on the current word during synthesis. In contrast, increasing \(k\) to 3 leads to a degradation in intelligibility, suggesting that larger lookahead is not always beneficial, as excessive future context may instead distract the model and result in less precise alignment.
In addition, UTMOS \cite{saeki22c_interspeech} indicates that setting \(k = 2\) results in more natural-sounding speech than \(k = 1\). To validate this observation, we conduct a preference test on 50 paired LibriTTS samples with 20 evaluators comparing the \(k = 1\) and \(k = 2\) variants. The results indicate that the \(k = 2\) variant is preferred in 65.9\% of the comparisons. Therefore, we adopt \(k = 2\) as the optimal trade-off for distillation and subsequent evaluations.

\vspace{-2mm}
\begin{table}[h!]
\caption{Ablation study of lookahead-causal masks (LCMs).}
\vspace{-3mm}
\centering
\small
\renewcommand{\arraystretch}{0.9}
\setlength{\aboverulesep}{0pt}
\setlength{\tabcolsep}{5pt}
\resizebox{1.0\columnwidth}{!}{%
\begin{tabular}{@{}lcccccc@{}}
\toprule
\textbf{Model Variant} 
& \textbf{CER $\downarrow$} 
& \textbf{WER $\downarrow$} 
& \textbf{Ins. $\downarrow$}
& \textbf{Del. $\downarrow$}
& \textbf{Sub. $\downarrow$}
& \textbf{SSIM $\uparrow$} \\
\midrule
S5-TTS
    & \textbf{2.12\%}
    & \textbf{3.49\%}
    & \textbf{0.66\%}
    & 1.01\%
    & \textbf{0.46\%}
    & \textbf{0.9328} \\
\midrule
\quad w/o both LCMs
    & 12.53\%
    & 14.20\%
    & 10.20\%
    & 1.00\%
    & 1.33\%
    & 0.9310 \\
\quad w/o enc. LCM
    & 33.04\%
    & 40.15\%
    & 26.54\%
    & 2.55\%
    & 3.96\%
    & 0.9281 \\
\quad w/o dec. LCM
    & 3.41\%
    & 4.92\%
    & 2.32\%
    & \textbf{0.36\%}
    & 0.73\%
    & 0.9323 \\
\bottomrule
\end{tabular}
}
\label{tab:ablation_lcm}
\vspace{-3mm}
\end{table}

\noindent\textbf{Ablation Study of Lookahead-Causal Masks:}
We investigate the impact of lookahead-causal masks (LCMs) on the same LibriTTS test utterances described earlier. As shown in Table~\ref{tab:ablation_lcm}, removing both LCMs leads to a substantial degradation in intelligibility. Notably, removing only the encoder LCM leads to an even higher WER than removing both LCMs, as the encoder is trained on full sequences, causing the decoder to depend on complete contextual information. During inference, the lack of future context disrupts this dependency, leading to increased errors. In comparison, removing both LCMs ensures consistency between training and inference, as the model treats each input with lookahead words as a complete sequence during inference, though overall quality degrades. Removing the decoder LCM has a smaller but also notable impact on WER. Overall, the results confirm that both encoder and decoder LCMs are essential for preserving S5-TTS performance under limited lookahead.

\noindent\textbf{Ablation Study of Distillation:}
We conduct an ablation study on the proposed distillation method IMSD by applying it to the \(k = 2\) variant of S5-TTS. As shown in Table~\ref{tab:cer_wer_ssim}, incorporating IMSD substantially improves intelligibility and naturalness over the naive S5-TTS, with WER reduced from 3.49\% to 2.65\% and UTMOS increased from 3.66 to 3.72 on LibriTTS. Since UltraChat is used for distillation, we further evaluate 200 unseen UltraChat test sentences using a high-quality proprietary female voice as reference. The distilled S5-TTS achieves the best intelligibility among the compared variants on this set as well. No notable change in SSIM is observed after distillation.

\begin{table}[h!]
\caption{Comparison with other AR and NAR TTS models.}
\vspace{-3mm}
\centering
\normalsize
\renewcommand{\arraystretch}{0.9}
\setlength{\aboverulesep}{0pt}
\setlength{\tabcolsep}{3pt}
\resizebox{1.0\columnwidth}{!}{%
\begin{tabular}{@{}cccccccc@{}}
\toprule
\textbf{Model} 
& \textbf{Params} 
& \textbf{Data (h)} 
& \textbf{WER $\downarrow$} 
& \textbf{UTMOS $\uparrow$} 
& \textbf{STOI $\uparrow$} 
& \textbf{PESQ $\uparrow$}
& \textbf{SSIM $\uparrow$} \\
\midrule
S5-TTS w/ IMSD
    & \textbf{160M} & \textbf{4.67K} & 2.65\%
    & 3.72
    & \textbf{0.179}
    & \textbf{1.075}
    & 0.9340 \\
\cmidrule(lr){1-8}
E2-TTS~\cite{eskimez2024e2}
    & 335M & 100K & 2.82\%
    & 3.65 & 0.137 & 1.071 & 0.9487 \\
FireRedTTS~\cite{guo2024fireredtts}
    & 400M & 248K & 4.70\%
    & 3.82 & 0.157 & 1.074 & 0.9229 \\
MaskGCT~\cite{wang2024maskgct}
    & 315M & 100K & \textbf{2.31\%}
    & 3.74 & 0.156 & 1.071 & \textbf{0.9495} \\
CosyVoice~\cite{du2024cosyvoice}
    & 300M & 170K & 2.46\%
    & \textbf{3.95} & 0.152 & 1.060 & 0.9117 \\
\bottomrule
\end{tabular}
}
\label{tab:tts_comparison_no_cer}
\vspace{-5mm}
\end{table}

\noindent\textbf{Comparison with Other TTS Models:} S5-TTS is primarily designed to provide a practical incremental synthesis approach for T5-based TTS models. For reference, we also compare the distilled S5-TTS with representative AR and NAR TTS models, including E2-TTS, MaskGCT, CosyVoice, and FireRedTTS, on the same LibriTTS test utterances described earlier. Notably, all compared baseline models are trained on more than 100K hours of speech, whereas the distilled S5-TTS is trained using only 4.67K hours in total. As Table 3 shows, S5-TTS outperforms all compared models in terms of STOI \cite{5495701} and PESQ~\cite{941023}. In addition, S5-TTS achieves lower WER than E2-TTS and FireRedTTS, and yields higher UTMOS than E2-TTS.

\noindent\textbf{Subjective Evaluation of Speech Quality:}
We conduct MOS evaluations on the LibriTTS and UltraChat testsets. For each model, 50 audio samples are rated by evaluators using a 5-point scale. For UltraChat, the same reference audio sample is used as described earlier. The MOS results with 95\% confidence intervals are reported in Table~\ref{tab:mos_and_performance}. Despite achieving comparable WER and SSIM, the naive S5-TTS shows a noticeable degradation in perceived quality compared to T5-TTS. In contrast, the distilled S5-TTS achieves MOS scores close to those of T5-TTS on both datasets, with gaps of only 0.04 on LibriTTS and 0.09 on UltraChat. This indicates that the proposed distillation method IMSD effectively restores the perceived speech quality and naturalness of S5-TTS under limited lookahead, achieving performance comparable to full-context T5-TTS.

\vspace{-2mm}
\begin{table}[h!]
\caption{MOS and Efficiency Metrics (latency in seconds).}
\vspace{-3mm}
\centering
\small
\setlength{\aboverulesep}{0pt}
\setlength{\tabcolsep}{6pt}
\resizebox{\columnwidth}{!}{%
\begin{tabular}{cccccc}
\toprule
\textbf{Eval Set} & \textbf{Model} & \textbf{MOS $\uparrow$} & \textbf{RTF $\downarrow$} & \textbf{FCL $\downarrow$} & \textbf{E2E $\downarrow$} \\
\midrule
\multirow{4}{*}{\shortstack{LibriTTS \\ \footnotesize (Unseen)}}
    & Ground Truth        & 3.81 $\pm$ 0.065 & --     & --     & --     \\
    & T5-TTS    & \textbf{3.75 $\pm$ 0.064} & 0.728 & 0.262 & 0.728 \\
    & S5-TTS    & 3.64 $\pm$ 0.067 & 0.616 & 0.181 & 0.354 \\
    & S5-TTS w/ IMSD  & \textbf{\textit{3.71 $\pm$ 0.062}} & \textbf{0.609} & \textbf{0.169} & \textbf{0.343} \\
\midrule
\multirow{3}{*}{\shortstack{UltraChat \\ \footnotesize (Unseen)}}
    & T5-TTS    & \textbf{4.21 $\pm$ 0.051} & 0.722 & 0.263 & 0.868 \\
    & S5-TTS    & 3.99 $\pm$ 0.058 & \textbf{0.615} & 0.186 & 0.365 \\
    & S5-TTS w/ IMSD  & \textbf{\textit{4.12 $\pm$ 0.054}} & 0.618 & \textbf{0.176} & \textbf{0.356} \\
\bottomrule
\end{tabular}
}
\label{tab:mos_and_performance}
\vspace{-3mm}
\end{table}

\noindent\textbf{Analysis of Model Efficiency:}
We compare S5-TTS with T5-TTS using naive PyTorch on the same B200 GPU used for training. For fairness, T5-TTS decoding is performed in a streaming manner with a fixed chunk size of 30 steps, closely matching the average step count of S5-TTS (30.2) measured on UltraChat. As shown in Table~\ref{tab:mos_and_performance}, we report the Real-Time Factor (RTF) and First-Chunk Latency (FCL) of the standalone TTS models. S5-TTS consistently achieves lower RTF and FCL than T5-TTS, and the distilled S5-TTS further reduces FCL.
We attribute this reduction to the improved naturalness introduced by distillation, which leads to fewer codec frames for the initial word. We further evaluate the end-to-end speech response latency (E2E) by integrating the TTS models with a Llama~3.3~70B \cite{grattafiori2024llama} LLM deployed via Ollama~\cite{ollama2026} at INT4 precision. In this setting, S5-TTS can begin synthesis after receiving the third word from the LLM (with \(k=2\)), whereas T5-TTS must wait for the complete sentence, resulting in substantially higher end-to-end latency.

\section{Conclusion}
\label{sec:conclusion}

We presented S5-TTS, a streaming text-to-speech model based on language modeling that demonstrates low-latency, word-by-word speech synthesis under limited lookahead. S5-TTS achieves speech quality comparable to full-context T5-TTS, supports zero-shot synthesis, and significantly reduces response latency in cascaded LLM-TTS systems. Experimental results underscore the feasibility of natural, truly streaming neural TTS and its potential for practical real-time conversational AI.

\section{Use of Generative AI Disclosure}
Generative AI tools were used solely for grammar checking and language polishing to improve the clarity of the manuscript.

\bibliographystyle{IEEEtran}
\bibliography{mybib}

\end{document}